# Expeditious Stochastic Calculation of Random-Phase Approximation Energies for Thousands of Electrons in 3 Dimensions

Daniel Neuhauser†, Eran Rabani‡ and Roi Baer♣

† Department of Chemistry and Biochemistry, University of California, Los Angeles CA 90095, USA.
‡ School of Chemistry, The Sackler Faculty of Exact Sciences, Tel Aviv University, Tel Aviv 69978, Israel.
♣ Fritz Haber Center for Molecular Dynamics, Institute of Chemistry, Hebrew University, Jerusalem 91904 Israel.

**ABSTRACT**: A fast method is developed for calculating the Random-Phase-Approximation (RPA) correlation energy for density functional theory. The correlation energy is given by a trace over a projected RPA response matrix and the trace is taken by a stochastic approach using random perturbation vectors. The method scales, at most, quadratically with the system size but in practice, due to self-averaging, requires less statistical sampling as the system grows and the performance is close to linear scaling. We demonstrate the method by calculating the RPA correlation energy for cadmium selenide and silicon nanocrystals with over 1500 electrons. In contrast to 2nd order Møller-Plesset correlation energies, we find that the RPA correlation energies per electron are largely independent on the nanocrystal size.

Local and semi-local correlation functionals of the Kohn-Sham (KS) density functional theory (DFT) fail to describe for long-range van der Waals interactions and other types of dynamical screening effects.[1,2] One route for overcoming these deficiencies is the RPA theory[1,3-6] based on the KS-DFT adiabatic connection formalism[7-9] in combination with the fluctuation dissipation theorem.[10] In recent years this approach, especially when combined with exact exchange, was used successfully for treating various ailments of KS-DFT in molecular and condensed matter systems[5,6,11-18].

The greatest hurdle facing widespread use of RPA in systems of interest is its exceedingly high computational cost. Several approaches have been developed [5,6,13,19,20] for reducing the naïve $O(\widetilde{N}^6)$ RPA scaling to $O(\widetilde{N}^4)$, ($\widetilde{N}$ is a measure of system size); however, this is still expensive. The problem is aggravated when plane-waves or real-space grids are used, suffering from the huge number of unoccupied states and the strong reliance of the RPA energy on the unoccupied energies.[21-23]

In the present letter, we develop a stochastic sampling method for estimating the RPA correlation energy. Related sampling techniques have been recently developed by us for estimating the rate of multiexciton generation in nanocrystals (NCs),[24] for a linear scaling calculation of the exchange energy,[25] and for overcoming the computational bottleneck in Moller-Plesset second order perturbation theory (MP2).[26]

RPA, applied on top of a grid or plane waves calculation, starts from the KS Hamiltonian $\widehat{H}_0$ which can be applied to any wave function in a linear scaling way.[27] For a closed shell system of $2N$ electrons the $N$ lowest energy eigenfunctions $\phi_i(r)$ of $\widehat{H}_0$ are occupied and $M - N$ are unoccupied.[28] The RPA correlation energy can be written as[5] $E_C^{RPA} = \frac{1}{2}\sum_{ia,\widetilde{\Omega}_{ia}>0}(\widetilde{\Omega}_{ia} - A_{ia,ia})$, where $\widetilde{\Omega}_{ia}$ are eigenvalues of "$\widetilde{L}$" defined by:

$$\widetilde{L}\begin{pmatrix}X\\Y\end{pmatrix}_{ak} = \begin{pmatrix}0 & B-A\\A+B & 0\end{pmatrix}\begin{pmatrix}X\\Y\end{pmatrix}_{ak} = i\widetilde{\Omega}_{ak}\begin{pmatrix}X\\Y\end{pmatrix}_{ak}, \quad (1)$$

and

$$A_{ka,jb} = 2W_{ka,jb} + \delta_{kj}\delta_{ab}\omega_{kj} \quad (2)$$

$$\widetilde{B}_{ka,jb} = 2W_{ka,jb},$$

where $W_{pq,st} = \lambda\int\frac{\phi_p(r)\phi_s(r')\phi_t(r')\phi_q(r)}{|r-r'|}d^3rd^3r'$ are the Coulomb integrals, $\lambda$ is a coupling strength parameter ($\lambda = 1$ is full-strength Coulomb coupling and $\lambda = 0$ is the non-interacting limit), and $\omega_{st} = \varepsilon_s - \varepsilon_t$ us the difference of eigenvalues of $\widehat{H}_0$. Usually, all $\widetilde{\Omega}_{ak}$ are real, however $\widetilde{L}$, although having pure imaginary eigenvalues, is a real matrix operating on real vectors $\begin{pmatrix}X\\Y\end{pmatrix}$. Note that $\widetilde{\Omega}_{ia}$ are also the eigenvalues of the matrix $\begin{pmatrix}A & B\\-B & -A\end{pmatrix}$ appearing in standard RPA treatments.[5]

An alternative formulation starts from the expression:[5]

$$E_C^{RPA} = \frac{1}{2}\left[\widetilde{R}(1) - \left(\widetilde{R}(\lambda) + \frac{d\widetilde{R}}{d\lambda}\right)_{\lambda=0}\right], \quad (3)$$

Where $\widetilde{R}(\lambda) \approx \frac{1}{2}\sum_{ia,\widetilde{\Omega}_{ia}>0}\widetilde{\Omega}_{ia}(\lambda)$. However, the calculation of $\widetilde{R}(\lambda)$ is still prohibitive for large systems because of the high cost of diagonalization of the $2N(M-N) \times 2N(M-N)$ $\widetilde{L}$ matrix (in grid representations $M$ and $N$ easily reach $10^6$ and $10^4$ respectively).

Our formulation is based on linear-response time-dependent Hartree approach.[29,30] $E_C^{RPA}$ is still given by Eq. (3) but $\widetilde{R}(\lambda)$ is replaced by the following trace:

$$R(\lambda) \equiv \text{tr}\left[\Omega^+\left(i\widehat{L}(\lambda)\right)\right]. \quad (4)$$

Here $\Omega^+(x) = \frac{1}{2}x\theta(x)$, where $\theta(x)$ is the Heaviside step function, which we approximate as $\theta(x) \approx \frac{1}{2}\text{erfc}(-\beta x)$; $\widehat{L}$ is a *linear operator* defined by:[14,30]

$$\widehat{L}\begin{pmatrix}\chi_k\\Y_k\end{pmatrix} = \begin{pmatrix}-(\widehat{H}_0 - \epsilon_k)Y_k\\v_H[\Delta\rho]\phi_k + (\widehat{H}_0 - \epsilon_k)\chi_k\end{pmatrix}. \quad (5)$$

$\chi_k(r)$ and $Y_k(r)$ are functions, originally describing the time-dependent Hartree response, but are used here as stochastic perturbations as detailed below. $v_H[\Delta\rho](r) = \lambda\int\frac{\Delta\rho(r')}{|r-r'|}d^3r'$ is the Hartree perturbation potential depending linearly on the

$\chi$'s via

$$\Delta\rho(\mathbf{r}) = 4\sum_{j=1}^{N}\phi_j(\mathbf{r})\chi_j(\mathbf{r}). \quad (6)$$

One can expand $\hat{L}$ in the basis of the KS orbitals and obtain its $2NM \times 2NM$ matrix $L_{pq,st}$ having $NM$ positive imaginary eigenvalues $i\Omega_{st}$ and an equivalent negative set. $\Omega_{st}$ can be divided into $N(M-N)$ "occupied-unoccupied" transitions $\Omega_{ka}$ and $N^2$ occupied-occupied transitions $\Omega_{kj}$. Obviously, the dimensions of $L$ and $\tilde{L}$ differ, as the latter describes only occupied-unoccupied transitions. Nonetheless, within the occupied-unoccupied space the matrices and eigenvalues are identical:[31]

$$L_{ka,jb} = \tilde{L}_{ka,jb}, \quad \Omega_{ka} = \tilde{\Omega}_{ka}. \quad (7)$$

$R(\lambda)$ in Eq. (4) is computed in two principal steps:

1) The trace is replaced by an average (denoted by curly brackets) over random perturbation vectors $\binom{\chi}{\Upsilon}$:

$$R(\lambda) = \left\{ \left\langle (\chi \quad \Upsilon) | \Omega^+(i\hat{L}(\lambda)) | \binom{\chi}{\Upsilon} \right\rangle \right\}. \quad (8)$$

At each grid point $\mathbf{r}_g$ we set $\chi_k(\mathbf{r}_g) = \frac{1}{h^{3/2}}\xi_g$, $\Upsilon_k(\mathbf{r}_g) = \frac{1}{h^{3/2}}\eta_{kg}$, where $\xi_{kg}$ and $\eta_{kg}$ are independent random variables with values 1 or $-1$, selected such that $\{\xi_{kg}, \xi_{k'g'}\} = \{\eta_{kg}, \eta_{k'g'}\} = \delta_{gg'}\delta_{kk'}$ and $\{\xi_{kg}, \eta_{k'g'}\} = 0$. Thus, one has $\{\langle\chi|\chi\rangle\} = \{\langle\Upsilon|\Upsilon\rangle\} = NN_g$ and $\{\langle(\chi \quad \Upsilon)|\binom{\chi}{\Upsilon}\rangle\} = 2NN_g$, where $N_g$ is the grid size.

2) The action of the operator $\Omega^+(i\hat{L})$ on the random $\binom{\chi}{\Upsilon}$ is performed using an iterative modified Chebyshev polynomial expansion approach, so that:

$$R(\lambda) = \sum_{m=0}^{n_c} c_m r_m, \quad (9)$$

where

$$r_m = \left\{ \left\langle (\chi \quad \Upsilon) | \binom{\chi_m}{\Upsilon_m} \right\rangle \right\} \quad (10)$$

are the modified Chebyshev residues, and $\binom{\chi_m}{\Upsilon_m}$ are calculated iteratively

$$\binom{\chi_{m+1}}{\Upsilon_{m+1}} = \frac{2}{\Delta}\hat{L}\binom{\chi_m}{\Upsilon_m} + \binom{\chi_{m-1}}{\Upsilon_{m-1}} \quad (m > 1), \quad (11)$$

with

$$\binom{\chi_0}{\Upsilon_0} = \binom{\chi}{\Upsilon}, \quad \binom{\chi_1}{\Upsilon_1} = \frac{1}{\Delta}\hat{L}\binom{\chi_0}{\Upsilon_0}. \quad (12)$$

Note that $\hat{L}$ is a real operator (Eq. (5)) so all calculations are done on real functions. $\Delta = \frac{1}{2}(l_{max} - l_{min})$ is half the eigenvalue range of $i\hat{L}$. The $c_m$ are numerical coefficients obtained as follows: First, prepare a series of length $4n_c$, $d_n = \Omega^+\left(\Delta\cos\left(\frac{\pi}{2n_c}n\right)\right)$, $n = 0,\dots,4n_c - 1$ and then set $c_m = \frac{\tilde{d}_m}{4n_c(1+\delta_{m0})}$ (for $m = 0,\dots,n_c$) where $\{\tilde{d}\}$ are the discrete Fourier transform of $\{d\}$. The series length $n_c$ is chosen large enough so that the sum in Eq. (9) converges, i.e. $|c_{n_c}|$ is smaller than a prescribed tolerance.

Table 1: Parameters for the CdSe nanocrystals NCs. Shown are the number of Cd ($N_{Cd}$) and Se ($N_{Se}$) atoms, electrons ($N_e$), NC diameter ($D$), the numerical effort involved in operating with $L$ in a perturbation vector $N_L = \frac{1}{2}N_e \times N_g$, where $N_g$ is the number of grid-points and the occupied-unoccupied energy gap $E_g$.

| $N_{Cd}$ | $N_{Se}$ | $N_e$ | D (nm) | $N_L$ | $E_g$(eV) |
|---|---|---|---|---|---|
| 20 | 19 | 152 | 1.4 | 2,490,368 | 3.8 |
| 83 | 81 | 648 | 2.1 | 35,831,808 | 2.9 |
| 151 | 147 | 1176 | 2.5 | 154,140,672 | 2.7 |

Table 2: Same as Table 1, but for hydrogen passivated silicon NCs.

| $N_{Si}$ | $N_H$ | $N_e$ | D (nm) | $N_L$ | $E_g$(eV) |
|---|---|---|---|---|---|
| 1 | 4 | 8 | | 6,912 | 10.7 |
| 35 | 36 | 176 | 1.3 | 2,883,584 | 3.9 |
| 87 | 76 | 424 | 1.6 | 23,445,504 | 3.2 |
| 353 | 196 | 1608 | 2.4 | 210,763,776 | 2.2 |

We rely on correlated sampling to reduce the statistical error in computing $E_C^{RPA}$: We compute $R$ for 3 values of $\lambda = 1, +\eta$ and $-\eta$ (with $\eta = 10^{-3}$) using the same random number seeds, and then the RPA energy is estimated by:

$$E_C^{RPA} = \frac{1}{2}\left[R(1) - \frac{1}{2}\sum_{\lambda=\pm\eta}\left(1 + \frac{1}{\lambda}\right)R(\lambda)\right]. \quad (13)$$

We now demonstrate the performance of the stochastic method by applying it to calculate the RPA correlation energies of spherical cadmium-selenide (CdSe) and hydrogen passivated silicon NCs, where the Hamiltonian $\hat{H}_0$ is constructed from a semiempirical pseudopotential model.[32,33] The $N$ occupied states of the NCs were obtained using the filter diagonalization technique[34] with the implementation described in Refs. [32,35]. We used $\beta = 30E_h^{-1}$ for approximating the step function $\Omega^+(x)$ and $n_c = 1024$ (see discussion of Figure 3), and $\Delta \approx 12E_h$, slightly larger than half the maximal eigenvalue range for both NCs. Various features of the NCs are summarized in Table 1 and 2.

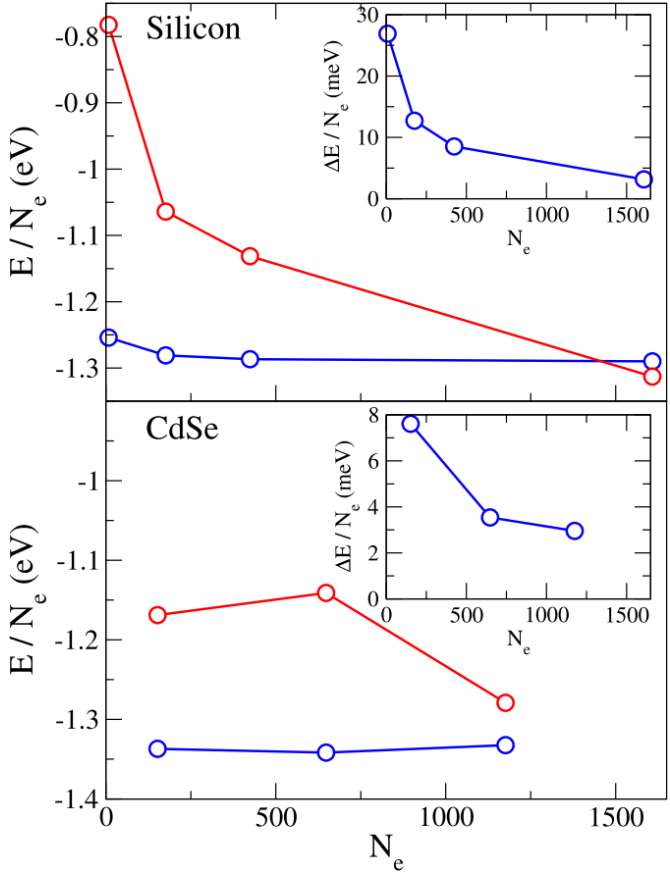

Figure 1: RPA (blue) and MP2 (red) correlation energies per electron vs. the number of electrons $N_e$, for silicon (top) and CdSe (bottom) NCs. Insets show the statistical errors in the RPA energies, normalized to 1000 stochastic iterations.

As a test, we compared the stochastic estimate and a full summation calculation of the RPA energy for the smallest silicon system on a $8 \times 8 \times 8$ point grid. We found that for 10,000 sampling iterations the stochastic estimate deviates by ~10meV from the full summation value. Such deviation is comparable to the Chebyshev truncation error at $n_c = 256$.

Figure 1 shows the RPA correlation energies for CdSe and silicon NCs up to $\approx 1600$ electrons along with a comparison to MP2 energies obtained using the Neuhauser-Rabani-Baer (NRB) method.[26] The RPA correlation energy depends weakly on NC size in contrast to that of MP2. This is because the NC gaps decrease with system size and MP2 energies are sensitive to small gaps (diverging for metals). The RPA energy of silicon is somewhat above that of CdSe, and is within the LDA bulk limit[21] range of $1 - 1.5$ eV.

The insets of Figure 1 show the corresponding statistical errors normalized to 1000 stochastic iterations. The errors decrease when the number of electrons in the system increases. This shows that the algorithm profits from statistical self-averaging. The statistical error of the CdSe NCs is approximately twice smaller than that for silicon despite having similar gaps for the same NC size. This suggests that the statistical errors are not trivially correlated with the gap.

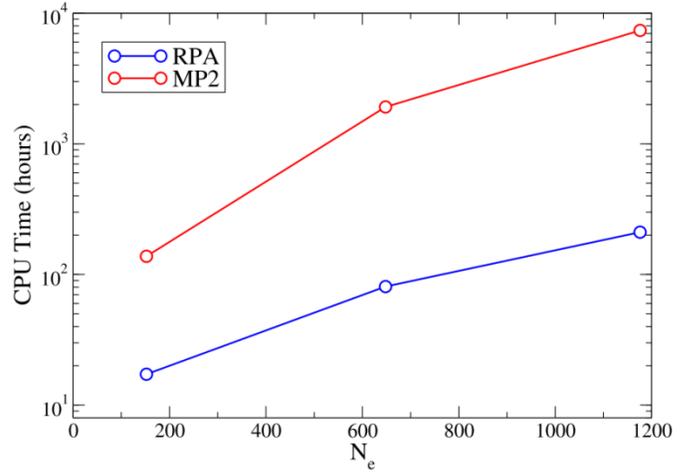

Figure 2: The CPU times for achieving a statistical error of $\approx 10$meV per electron for the RPA and MP2 calculations of CdSe NCs.

Figure 2 shows the total CPU time for calculations that yield a statistical error of $\approx 10$meV per electron. The method scales, at most, quadratically with system size but in practice, due to self-averaging, requires considerably less statistical sampling as the system grows and the resulting performance is close to *linear scaling*. Furthermore, in comparison, the RPA CPU time for the same statistical error is an order of magnitude smaller than the CPU time required for the MP2 calculations. Regarding memory requirements, for the RPA is scales quadratically with system size (17GB for the largest silicon NC) and linearly for MP2.

For some systems, mainly with small gaps, the RPA matrix $iL$ may include several imaginary eigenvalues, because electron-hole interactions are not properly screened and become larger than the quasiparticle gap. In this case the Chebyshev interpolation method, which only samples the real axis, can become unstable. This is shown in the upper panel of Figure 3 for the $Si_{353}H_{196}$ NC where we plot the RPA energy estimate as function of the length of the Chebyshev expansion. The results (dashed line) clearly diverge as the Chebyshev expansion length grows. For CdSe NCs (results shown in the lower panel of Figure 3) this instability is absent.

When the Chebyshev expansion diverges we develop a different interpolation polynomial scheme which samples interpolation points in both the real and imaginary axes, based on the Newton interpolation.[36] This method is effective, as seen in the solid line corresponding to the $Si_{353}H_{196}$ NC, however it requires 20% times more memory and 4 times as much CPU time. Details of this approach will be described in future publications.

Summarizing, we have presented a new stochastic approach for calculating RPA energies for large electronic systems of exceptional size. The method scales formally as $O(\tilde{N}^2)$ in terms of memory and CPU time but due to self-averaging has a near-linear scaling performance. We calculated the RPA correlation energy for CdSe and silicon NCs up to diameters of 2.5nm with over 1500 electrons. The stochastic approach developed here bloats, by orders of magnitude, the size of systems that can be treated using RPA theory.

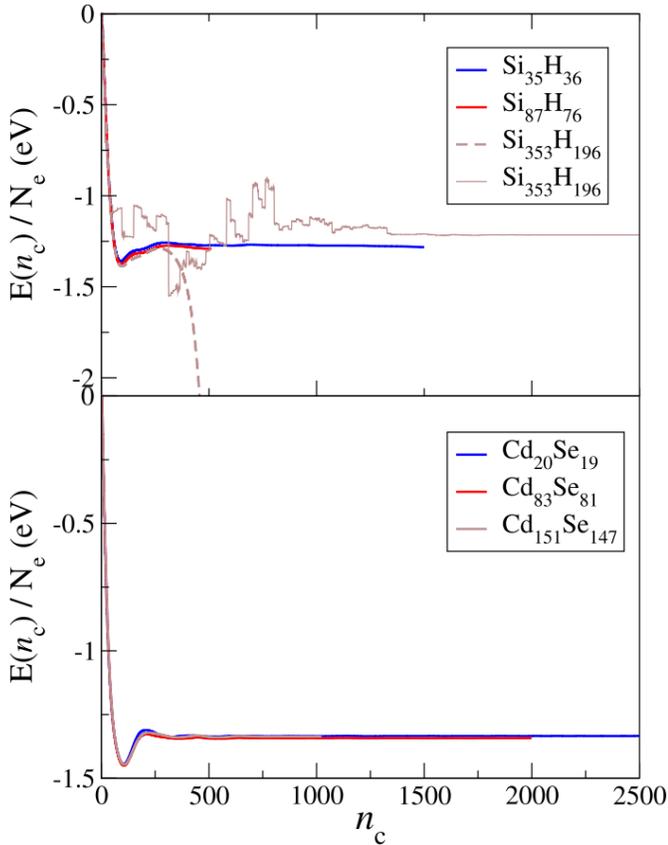

Figure 3: The RPA correlation energy as function of the length of the Chebyshev interpolation polynomial for silicon (top) and CdSe (bottom) NCs. For the largest silicon NC the dashed line shows divergence of the Chebyshev scheme. The solid line for this case uses a Newton interpolation with sampling points in the complex plane.

DN was supported by the DOE MEEM center, award DE-SC0001342. RB was supported by the US-Israel Binational Foundation (BSF). RB and ER gratefully thank the Israel Science Foundation, grant numbers 1020/10 and 611/11, respectively.

1. J. P. Perdew and K. Schmidt, in *Density functional theory and its application to materials : Antwerp, Belgium, 8-10 June 2000*, edited by V. E. Van Doren, C. Van Alsenoy and P. Geerlings (American Institute of Physics, Melville, N.Y., 2001).
2. N. Marom, A. Tkatchenko, M. Rossi, V. V. Gobre, O. Hod, M. Scheffler, and L. Kronik, J. Chem. Theor. Comp **7**, 3944 (2011).
3. B. G. Janesko, T. M. Henderson, and G. E. Scuseria, J. Chem. Phys. **131**, 034110 (2009).
4. M. Fuchs and X. Gonze, Phys. Rev. B **65**, 235109 (2002).
5. H. Eshuis, J. E. Bates, and F. Furche, Theor. Chem. Acc. **131** (2012).
6. X. G. Ren, P. Rinke, C. Joas, and M. Scheffler, J. Mater. Sci. **47**, 7447 (2012).
7. D. C. Langreth and J. P. Perdew, Sol. Stat. Comm. **17**, 1425 (1975).
8. O. Gunnarsson and B. I. Lundqvist, Phys. Rev. B **13**, 4274 (1976).
9. D. C. Langreth and J. P. Perdew, Phys. Rev. B **15**, 2884 (1977).
10. H. B. Callen and T. A. Welton, Phys. Rev. **83**, 34 (1951).
11. Y. Andersson, D. C. Langreth, and B. I. Lundqvist, Phys. Rev. Lett. **76**, 102 (1996).
12. J. F. Dobson and J. Wang, Phys. Rev. Lett. **82**, 2123 (1999).
13. J. Paier, X. Ren, P. Rinke, G. E. Scuseria, A. Gruneis, G. Kresse, and M. Scheffler, New J. Phys. **14**, 043002 (2012).
14. R. Baer and D. Neuhauser, J. Chem. Phys. **121**, 9803 (2004).
15. G. E. Scuseria, T. M. Henderson, and D. C. Sorensen, J. Chem. Phys. **129**, 231101 (2008).
16. J. G. Angyan, R. F. Liu, J. Toulouse, and G. Jansen, J. Chem. Theor. Comp **7**, 3116 (2011).
17. J. Paier, B. G. Janesko, T. M. Henderson, G. E. Scuseria, A. Gruneis, and G. Kresse, J. Chem. Phys. **132**, 094103 (2010).
18. L. Schimka, J. Harl, A. Stroppa, A. Gruneis, M. Marsman, F. Mittendorfer, and G. Kresse, Nature Materials **9**, 741 (2010).
19. D. Rocca, Z. Bai, R.-C. Li, and G. Galli, J. Chem. Phys. **136**, 034111 (2012).
20. X. G. Ren, P. Rinke, and M. Scheffler, Phys. Rev. B **80**, 045402 (2009).
21. H. V. Nguyen and S. de Gironcoli, Phys. Rev. B **79**, 205114 (2009).
22. J. Harl and G. Kresse, Phys. Rev. B **77**, 045136 (2008).
23. D. Lu, H.-V. Nguyen, and G. Galli, J. Chem. Phys. **133**, 154110 (2010).
24. R. Baer and E. Rabani, Nano Lett. **12**, 2123 (2012).
25. R. Baer and D. Neuhauser, J. Chem. Phys. **137**, 051103 (2012).
26. D. Neuhauser, E. Rabani, and R. Baer, J. Chem. Theor. Comp (2012).
27. We limit the present study to KS DFT or to DFT with short range hybrid functionals.
28. We denote occupied states by indices $i,j,k,l=1,..,N$ and unoccupied states by $a,b,c,d=N+1,…,M$; the indices $p,q,s,t=1,…,M$ denote "unrestricted" states.
29. D. Neuhauser and R. Baer, J. Chem. Phys. **123**, 204105 (2005).
30. S. Baroni, S. de Gironcoli, A. Dal Corso, and P. Giannozzi, Rev. Mod. Phys. **73**, 515 (2001).
31. The occupied-occupied eigenvalues, although affecting R will not affect the RPA correlation energy calculated by Eq. (3). A proof will be given in a future publication.
32. E. Rabani, B. Hetenyi, B. J. Berne, and L. E. Brus, J. Chem. Phys. **110**, 5355 (1999).
33. L. W. Wang and A. Zunger, J. Phys. Chem. **98**, 2158 (1994).
34. M. R. Wall and D. Neuhauser, J. Chem. Phys. **102**, 8011 (1995).
35. S. Toledo and E. Rabani, J. Comp. Phys. **180**, 256 (2002).
36. R. Kosloff, Ann. Rev. Phys. Chem. **45**, 145 (1994).

4